# A First Step for Expansion X-Ray Microscopy: Achieving Contrast in Expanded Tissues Sufficient to Reveal Cell Bodies

Logan Thrasher Collins[1,*]

[1]Washington University in St. Louis, Department of Biomedical Engineering
[*]Corresponding Author, clogan@wustl.edu

**Abstract:** Existing methods in nanoscale connectomics are at present too slow to map entire mammalian brains. As an emerging approach, expansion microscopy (ExM) has enormous promise, yet it still suffers from throughput limitations. Mapping the human brain and even mapping nonhuman primate brains therefore remain distant goals. While ExM increases effective resolution linearly, it enlarges tissue volume cubically, which dramatically increases imaging time. As a rapid tomographic technique, X-ray microscopy has potential for drastically speeding up large-volume connectomics. But to the best of my knowledge, no group has so far imaged cellular features within expanded tissue using X-ray microscopy. I herein present an early-stage report featuring the first demonstration of X-ray microscopy reconstruction of cell bodies within expanded tissue. This was achieved by combining a modified enzymatic Unclearing technique with a metallic gold stain and imaging using a laboratory X-ray microscope. I emphasize that a great deal of work remains to develop "expansion X-ray microscopy" (ExXRM) to the point where it can be useful for connectomics since the current iteration of ExXRM only resolves cell bodies and not neurites due to extensive off-target staining. Additionally, the current method must be modified to accommodate for the challenges of synchrotron X-ray microscopy, a vastly speedier approach than laboratory X-ray microscopy. Nonetheless, achieving X-ray contrast in expanded tissues represents a significant first step towards realizing ExXRM as a connectomics imaging modality.

**Introduction:**

Connectomics focuses on a longstanding goal in neuroscience: mapping the synaptic connections and morphological structures found in biological nervous systems. To achieve this, tools with the resolution needed to visualize structures like fine neurites (<100 nm diameter) and synapses (with ~20 nm synaptic clefts) are necessary. Electron microscopy (EM)[1,2] and fluorescence-based expansion microscopy (ExM)[3,4] are the primary methods used for connectomic imaging. Isotropic voxels of around 10-20 nm or non-isotropic voxels of around 10-45 nm are necessary to even have potential for accurate segmentation of neuronal tissue volumes.[1,5–7] Because of this, imaging entire mammalian brains will take prohibitively long periods of time with existing and near-future technologies.[5]

Synchrotron X-ray microscopy has promise for rapid imaging of brain tissue for connectomics applications. Specialized forms of synchrotron X-ray microscopy including X-ray holographic nano-tomography (XNH)[8–10] and ptychographic X-ray computed tomography (PXCT)[11] have already been implemented in the context of brain tissue imaging. However, these methods are still relatively slow and have only been applied in small (submillimeter) samples. Synchrotron full-field phase-contrast X-ray microtomography (XRM) offers dramatically faster imaging speeds (up to ~0.5 mm$^3$ per minute with current systems), yet is typically limited to isotropic voxel sizes of ~0.3 µm.[12,13] Imaging speed may increase further with development of larger detectors (see Supplementary Information after references) **(Figure S1)**.

Expansion microscopy involves infusing tissues with a swellable hydrogel and adding distilled water to trigger physical enlargement.[14,15] This revolutionary method has been applied in hundreds of laboratories across the world for a variety of applications.[16–19] ExM increases the effective resolution of imaging systems by enlarging tissues. If ExM were to be successfully combined with XRM, it would *theoretically* convert 0.3 µm voxels to effective 30 nm voxels (10× expansion) or effective 15 nm voxels (20× expansion). XRM's rapid imaging speed has potential to compensate for the increased imaging time resulting from the volumetric enlargement. However, ExM to date has mainly been used with fluorescence microscopy, where the diffuse illumination from fluorophores compensates for the "gaps" in the tissue which exist after expansion. For this reason, a new way of "filling in the gaps" is needed to properly implement ExM with XRM (ExXRM).

I herein showcase a first demonstration of ExXRM where expanded cell bodies are discernable. To the best of my knowledge, this represents the first time biological features of any kind have been successfully imaged using XRM in expanded tissue. To accomplish this, an ~18× expanded mouse cortex sample was prepared via pan-ExM[20] and signal amplified using enzymatic Unclearing for diaminobenzidine (DAB) deposition and a NanoProbes GoldEnhance kit for gold deposition.[21] The sample was then imaged using an Xradia 620 Versa laboratory X-ray microscope. The final expansion factor during X-ray imaging was ~8×, which increased the size of the cell bodies from ~10 µm to ~80 µm and made them visible despite a relatively large isotropic voxel size of 2.95 µm. This proof-of-concept takes an initial step towards showing how ExXRM may possess utility as a connectomics imaging technique.

Although laboratory X-ray microscopes are more easily accessible than synchrotron beamlines, there exist important differences between these two imaging modalities. Notably, synchrotrons produce X-rays at much higher flux than laboratory X-ray microscopes. While this increases imaging speed, the powerful radiation would certainly destroy the hydrogels used for ExM. Because of this, cryogenic imaging conditions will be crucial at synchrotron sources. Radioprotective reagents incorporated into the hydrogels may also help combat degradation of frozen expanded tissues.[11] But unlike laboratory X-ray microscopes, synchrotrons can achieve high resolution (and small voxel size) at long working distances,[22,23] which may facilitate imaging of large expanded samples in the future. The high brightness and coherence of synchrotron X-rays also greatly improves image quality and resolution compared to laboratory X-ray microscopes.[23,24] So, the successful identification of cell bodies here with a laboratory X-ray microscope system bodes positively for resolution of finer subcellular features in the future.

**Results:**

Through a contracted service by Panluminate Inc., mouse cortex samples were first expanded using pan-ExM, a type of iterative ExM which achieves expansion factors of up to 24-fold.[25] In this case, the expansion factor after pan-ExM was 18×. Panluminate pan-labeled the expanded tissue at primary amines with biotin, stained with Streptavidin-HRP (horseradish peroxidase), and applied a DAB solution. HRP's enzymatic activity facilitated deposition of DAB polymer in the proximity of labeled sites. This process of enzymatic DAB deposition, known as Unclearing, has previously been described by M'Saad et al.[21] Panluminate sectioned the expanded tissue gels to 200 µm thickness and re-embedded the samples in a secondary non-expanding hydrogel matrix to stabilize them and mitigate shrinkage. Next, Panluminate employed the GoldEnhance[TM] LM (light microscopy optimized) kit from NanoProbes to stain one of the two DAB-labeled samples. This kit is typically employed with colloidal nanogold clusters which serve as a substrate for catalytic deposition of gold ions, enlarging the clusters to much greater sizes (reportedly up to 40 µm). However, in this study Panluminate used DAB as the substrate for catalytic deposition of gold ions as based on the previous Unclearing work by M'Saad et al.[21] I thus sought to obtain gold-stained material *in situ* of sufficient absorption within the expanded tissue to achieve contrast with a laboratory X-ray microscope.

Before proceeding to X-ray imaging, brightfield microscopy was utilized to visualize the DAB only control sample **(Figure 1A)** and DAB-gold sample **(Figure 1B, Figure S2)**. Cell bodies and some partially visible filamentous structures were seen in both samples. Because of the gold deposition, the DAB-gold sample appeared much darker than the DAB-only sample. However, this coincided with evidence of significant off-target gold staining. Additionally, the GoldEnhance[TM] LM reagents did not fully diffuse throughout the entire expanded tissue volume, so only the edge portion of the sample received the gold labeling. Further work is needed to overcome these issues. It should be noted that, since the DAB-gold sample underwent partial shrinkage to a final expansion factor of ~8× after gold deposition, the cell body diameters were smaller than those in the DAB-only sample. After embedding the samples into 2 mm plastic tubes with agarose, a photograph was taken to illustrate their gross morphology **(Figure 1C)**.

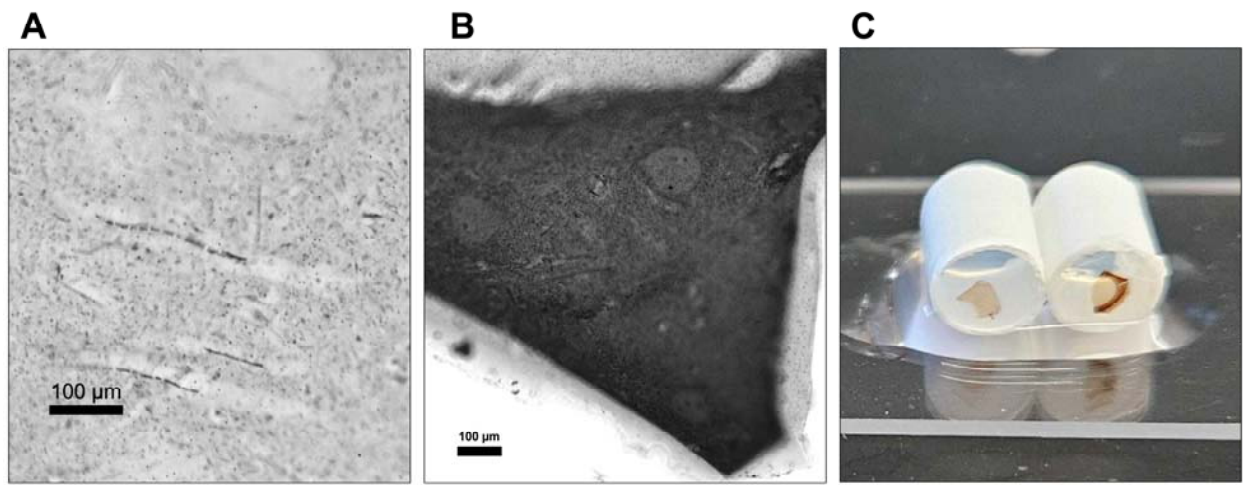

**Figure 1 (A)** Black-and-white brightfield microscopy image of expanded brain tissue Uncleared with DAB only (10× objective lens). **(B)** Black-and-white brightfield microscopy image of expanded brain tissue Uncleared with DAB and treated with the GoldEnhance[TM] LM kit (10× objective lens). **(C)** Photograph of the DAB-only (left) and DAB-gold (right) samples after agarose embedding inside of 2 mm plastic tubes.

Panluminate shipped the samples to the UTCT core facility of the University of Texas at Austin for XRM imaging. There, a Zeiss Xradia 620 Versa laboratory X-ray microscope was utilized to image the DAB-gold sample. To minimize desiccation, the 2 mm tube containing the sample was placed within a small vial containing PBS during imaging. Because the achievable voxel size of laboratory X-ray microscopes (but not synchrotron XRM systems) is relatively large when the detector and source are positioned at greater distance from the sample itself, the presence of the vial limited us to an isotropic voxel size of 2.95 µm.

XRM reconstruction revealed visible cell bodies in the sample **(Figure 2A-D)**. Neurites were unfortunately not decipherable in the 3D images, probably because the extensive off-target staining obscured smaller features. Interestingly, the cell bodies appeared visible by virtue of lower signal than the surrounding tissue, indicating that the extracellular space may have undergone gold enhancement more efficiently than the cytosolic compartment. Better-targeted signal amplification strategies will be crucial in future iterations of ExXRM technology.

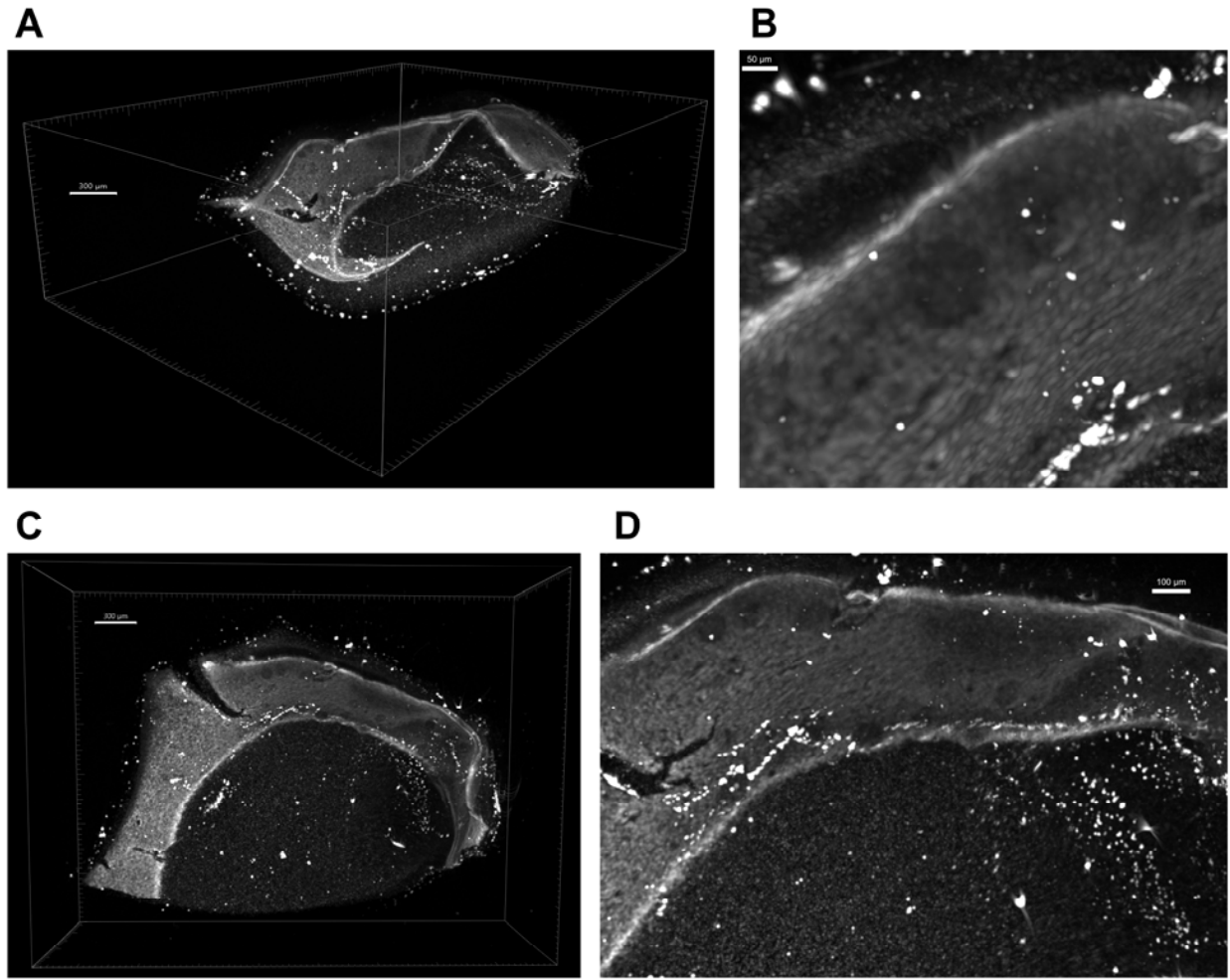

**Figure 2 (A)** Perspective view of the 3D XRM reconstruction of the expanded DAB-gold brain tissue sample. Cell bodies are visible as darker spots within the expanded tissue. The region where the gold reagents did not diffuse is visible as a grainy space without discernable morphological features. **(B)** Close up top view of the 3D XRM reconstruction of the expanded DAB-gold brain tissue sample showing round cell bodies of ~80 µm diameter. **(C)** Top view of the expanded DAB-gold brain tissue sample shows the same gross morphology as that seen in the earlier photograph of the sample. Cell body spots are visible. **(D)** Medium-scale close up top view of a portion of the expanded tissue containing several cell bodies.

**Discussion:**

The results of this study demonstrate that it is possible to leverage XRM to reconstruct features in expanded soft tissue. To the best of my knowledge, this is the first time such a result has been accomplished. However, important challenges remain before ExXRM can be utilized for connectomics. Firstly, the strategy of DAB Unclearing followed by GoldEnhance™ LM does not produce sufficiently targeted staining to delineate small cellular features (e.g. dendrites, axons, and synapses) by XRM or even by brightfield microscopy. Indeed, the visibility of cell bodies herein seemed to have arisen as a "negative space" effect due to less intense staining in the cytosol relative to the rest of the tissue. It should be noted that a similar effect might be exploited in the future by deliberately staining the extracellular spaces and not the intracellular compartments.[26] However, such an extracellular stain would likely need to be performed pre-expansion unless molecules unique to the extracellular space can be reliably targeted. As another possible future strategy, a mouse line (and perhaps eventually nonhuman primate line) genetically modified to express intracellularly localized or even membrane-localized SpyTag peptides[27] or similar[28] might allow much more targeted staining of the intracellular compartment or membrane. After expansion, localized SpyTag could be stained with SpyCatcher-HRP to facilitate DAB Unclearing and gold deposition. Alternatively, SpyCatcher linked to dendrimers decorated by gold nanoclusters may produce more controlled gold staining compared to enzymatically deposited DAB polymer. These approaches may improve the on-target/off-target ratio of GoldEnhance(TM) LM, revealing fine neuronal features like dendrites and axons.

As mentioned earlier, differences between laboratory XRM and synchrotron XRM present challenges for applying ExXRM to connectomics. Indeed, synchrotron X-ray beams can cause bubble formation in hydrated samples, destroying tissue structures. Although imaging at cryogenic temperature and infusing samples with radioprotective reagents[11] might prevent bubble formation and minimize radiation damage, this remains an unproven facet of the ExXRM technology. Additionally, metals like gold strongly absorb X-rays, which may exacerbate heat damage in expanded tissue even with cryopreservation. As such, it remains to be seen whether gold stains will translate from laboratory XRM to synchrotron XRM, or if signal enhancement with a different contrast agent would provide better outcomes. That said, there exist examples of evidence for the feasibility of using gold in a synchrotron context.[29–31] Of note, Depannemaecker et al. succeeded in synchrotron imaging of mouse brain tissue after staining neurons with both silver (via Golgi's method) and with 20 nm gold nanoparticles conjugated to anti-NeuN antibodies.[29] Future development of ExXRM metal-based stains may necessitate investigation of a range of concentrations of metals in expanded tissues to find the optimal level to achieve both strong contrast and sample stability.

Since the GoldEnhance(TM) LM kit used in this study was limited to a diffusion depth of a few hundred micrometers into the expanded tissue, ways of improving tissue penetration will be important when dealing with large brain volumes. Stochastic electrotransport, a method which leverages electric fields to push molecules through cleared tissues, may greatly enhance the penetration of the metallic stain reagents into expanded tissues.[32] Alternative hydrogel chemistries or modified metallic stain reagents may also be leveraged to improve diffusion. These strategies may facilitate gold enhancement throughout much larger tissue volumes.

This study provides a proof-of-concept demonstration showing that it is possible to use signal amplification methods to achieve sufficient contrast in expanded tissues to visualize cellular features. Based on the results of this investigation, I suggest the next step should be to develop a metallic signal amplification method with greatly improved on-target staining. This should reveal finer details such as dendrites and axons within expanded samples. Methods for improving penetration depth should be developed as well. From there, cryopreservation and other stabilization methods should be explored to enable synchrotron imaging. If the challenges thus far described can be overcome, ExXRM could greatly improve the throughput of connectomic imaging, possibly even providing a foundation for human whole-brain connectomes.

## Methods:

*Tissue Preparation and Expansion*

Brain tissue expansion was performed by Panluminate Inc. through a contracted service. Certain steps in the expansion workflow were performed using proprietary methods developed and optimized by Panluminate Inc. and are not disclosed in detail. Tissue expansion was performed as previously described in the pan-ExM protocol by M'Saad et al.[25] Briefly, mouse brain tissue was fixed via transcardial perfusion with a solution of acrylamide and formaldehyde and post-fixed overnight in an identical solution. Sections were subsequently sectioned to 70 µm, embedded in the initial expansion gel and placed in Milli-Q water until fully expanded. Gels were then re-embedded in an additional expanding gel. The gel was placed into Milli-Q water once again until it reached an expansion factor of about 18-fold.

*DAB-Based Unclearing*

Following expansion, tissue sections were washed three times in 1× PBS and incubated overnight at 4°C in a 50 µM working solution of EZ-Link™ NHS-PEG4-Biotin, No-Weigh™ Format (ThermoFisher, Cat. 26113). The following day, biotinylated samples were incubated overnight on a rocker with Pierce™ ExcelAid™ Streptavidin-HRP (ThermoFisher, Cat. N100) at a concentration of 2 µg/mL in 1× PBS. After HRP binding, samples were developed using a DAB substrate solution prepared according to the manufacturer's kit instructions, with development monitored at approximately 20, 40, and 60 minutes and stopped once adequate signal intensity was reached. After DAB development, samples were re-embedded into a non-expanding hydrogel matrix to stabilize the expanded structure and retain the DAB reaction product. This provided mechanical stability for subsequent gold-enhancement steps and prevented excessive gel shrinkage.

*Gold Enhancement*

Re-embedded tissue sections were sectioned on a vibratome to a thickness of 200 µm and treated with the GoldEnhance™ LM (Light Microscopy) Kit (Nanoprobes, Cat. 2113) to selectively deposit metallic gold onto the DAB reaction product. Enhancement was performed according to the manufacturer's instructions, and development was monitored for up to 1 hour or until adequate signal intensity was achieved.

*X-Ray Microtomography Imaging*

The DAB-gold sample was embedded in agarose inside of a 2 mm cylindrical plastic tube (acrylonitrile butadiene styrene). The sample was then shipped to the University of Texas High-Resolution X-ray CT Facility and scanned on a Zeiss 620 Versa X-ray microscope. The 2 mm tube was mounted in a vial and immersed in PBS to prevent desiccation. The X-ray source was set to 50 keV/4.5W, and 6001 unbinned projections were acquired on the flat panel detector over 360 degrees of rotation, with 5 samples averaged per projection at 0.2 s acquisition time. An air filter and multiple averaged reference was employed, with a secondary reference acquired with an LE5 filter. The source position of -10.4 mm and the detector position of 254.1 mm resulted in 2.95 µm isotropic voxels. The data were reconstructed using Zeiss software, with a center shift of -1.924 and byte scaling of [-0.01, 0.5].

**Acknowledgements:** Sample preparation was performed using a contracted service from Panluminate Inc. where Jonathan Gulcicek carried out the experimental procedures and Ons M'Saad offered valuable guidance. X-ray microtomography imaging was performed by Jessica A. Maisano at the University of Texas High-Resolution X-ray CT Facility.

**Funding:** This project was funded entirely through an AI Safety Grant of $20,000 from the Foresight Institute, a nonprofit organization dedicated to advancing frontier technologies and helping build a bright future.

**References:**
1. Xu, C. S., Pang, S., Hayworth, K. J. & Hess, H. F. Enabling FIB-SEM Systems for Large Volume Connectomics and Cell Biology. *bioRxiv* 852863 (2019) doi:10.1101/852863.
2. Motta, A., Schurr, M., Staffler, B. & Helmstaedter, M. Big data in nanoscale connectomics, and the greed for training labels. *Curr. Opin. Neurobiol.* **55**, 180–187 (2019).
3. Tavakoli, M. R. *et al.* Light-microscopy based dense connectomic reconstruction of mammalian brain tissue. *bioRxiv* 2024.03.01.582884 (2024) doi:10.1101/2024.03.01.582884.
4. Park, S. Y. *et al.* Combinatorial protein barcodes enable self-correcting neuron tracing with nanoscale molecular context. *bioRxiv* 2025.09.26.678648 (2025) doi:10.1101/2025.09.26.678648.
5. Collins, L. T., Huffman, T. & Koene, R. Comparative prospects of imaging methods for whole-brain mammalian connectomics. *Cell Reports Methods* **5**, (2025).
6. Motta, A. *et al.* Dense connectomic reconstruction in layer 4 of the somatosensory cortex. *Science (80-. ).* **366**, eaay3134 (2019).
7. Schmidt, H. *et al.* Axonal synapse sorting in medial entorhinal cortex. *Nature* **549**, 469–475 (2017).
8. Kuan, A. T. *et al.* Dense neuronal reconstruction through X-ray holographic nano-tomography. *Nat. Neurosci.* **23**, 1637–1643 (2020).
9. Livingstone, J. *et al.* Scaling up X-ray holographic nanotomography for neuronal tissue imaging. *Biomed. Opt. Express* **16**, 2047–2060 (2025).
10. Nathansen, A. *et al.* Cell nuclei segmentation in mm-scale x-ray holographic nanotomography images of mouse brain tissue. in *Proc.SPIE* vol. 13152 131521B (2024).
11. Bosch, C. *et al.* Nondestructive X-ray tomography of brain tissue ultrastructure. *Nat. Methods* (2025) doi:10.1038/s41592-025-02891-0.
12. Stampfl, A. P. J. *et al.* SYNAPSE: An international roadmap to large brain imaging. *Phys. Rep.* **999**, 1–60 (2023).
13. Chen, H. H. *et al.* High-resolution fast-tomography brain-imaging beamline at the Taiwan Photon Source. *J. Synchrotron Radiat.* **28**, 1662–1668 (2021).
14. Chang, J.-B. *et al.* Iterative expansion microscopy. *Nat. Methods* **14**, 593–599 (2017).
15. Chen, F., Tillberg, P. W. & Boyden, E. S. Expansion microscopy. *Science (80-. ).* **347**, 543 LP – 548 (2015).

16. Sahabandu, N. *et al.* Expansion microscopy for the analysis of centrioles and cilia. *J. Microsc.* **276**, 145–159 (2019).
17. Gao, R. *et al.* Cortical column and whole-brain imaging with molecular contrast and nanoscale resolution. *Science (80-. ).* **363**, eaau8302 (2019).
18. Gambarotto, D. *et al.* Imaging cellular ultrastructures using expansion microscopy (U-ExM). *Nat. Methods* **16**, 71–74 (2019).
19. Halpern, A. R., Alas, G. C. M., Chozinski, T. J., Paredez, A. R. & Vaughan, J. C. Hybrid Structured Illumination Expansion Microscopy Reveals Microbial Cytoskeleton Organization. *ACS Nano* **11**, 12677–12686 (2017).
20. M'Saad, O. *et al.* All-optical visualization of specific molecules in the ultrastructural context of brain tissue. *bioRxiv* 2022.04.04.486901 (2022) doi:10.1101/2022.04.04.486901.
21. M'Saad, O., Shribak, M. & Bewersdorf, J. Unclearing Microscopy. *bioRxiv* 2022.11.29.518361 (2022) doi:10.1101/2022.11.29.518361.
22. Walsh, C. L. *et al.* Imaging intact human organs with local resolution of cellular structures using hierarchical phase-contrast tomography. *Nat. Methods* **18**, 1532–1541 (2021).
23. Albers, J., Svetlove, A. & Duke, E. Synchrotron X-ray imaging of soft biological tissues – principles, applications and future prospects. *J. Cell Sci.* **137**, jcs261953 (2024).
24. Bosch, C. *et al.* Functional and multiscale 3D structural investigation of brain tissue through correlative in vivo physiology, synchrotron microtomography and volume electron microscopy. *Nat. Commun.* **13**, 2923 (2022).
25. M'Saad, O. *et al.* All-optical visualization of specific molecules in the ultrastructural context of brain tissue. *Nat. Biotechnol.* (2025) doi:10.1038/s41587-025-02905-4.
26. Michalska, J. M. *et al.* Imaging brain tissue architecture across millimeter to nanometer scales. *Nat. Biotechnol.* **42**, 1051–1064 (2024).
27. Keeble, A. H. *et al.* Approaching infinite affinity through engineering of peptide–protein interaction. *Proc. Natl. Acad. Sci.* **116**, 26523 LP – 26533 (2019).
28. Keeble, A. H. *et al.* DogCatcher allows loop-friendly protein-protein ligation. *Cell Chem. Biol.* **29**, 339-350.e10 (2022).
29. Depannemaecker, D. *et al.* Gold Nanoparticles for X-ray Microtomography of Neurons. *ACS Chem. Neurosci.* **10**, 3404–3408 (2019).
30. Schultke, E. *et al.* Single-cell resolution in high-resolution synchrotron X-ray CT imaging with gold nanoparticles. *J. Synchrotron Radiat.* **21**, 242–250 (2014).
31. Liu, T., Kempson, I., de Jonge, M., Howard, D. L. & Thierry, B. Quantitative synchrotron X-ray fluorescence study of the penetration of transferrin-conjugated gold nanoparticles inside model tumour tissues. *Nanoscale* **6**, 9774–9782 (2014).
32. Kim, S.-Y. *et al.* Stochastic electrotransport selectively enhances the transport of highly electromobile molecules. *Proc. Natl. Acad. Sci.* **112**, E6274–E6283 (2015).
33. Yakovlev, M. A. *et al.* A wide-field micro-computed tomography detector: micron resolution at half-centimetre scale. *J. Synchrotron Radiat.* **29**, 505–514 (2022).

**Supplemental Information:**

*How fast can synchrotron XRM image tissues?*

Synchrotron XRM beamlines have not yet been optimized for extremely large-scale imaging projects and thus still utilize relatively small detectors. Larger detectors[33] (which maintain ~0.3 µm voxel size) would dramatically increase the imaging rate, unlocking the full potential of ExXRM as an imaging modality **(Figure S1A-B)**. The theoretical imaging rate of an XRM system in mm$^3$/minute can be calculated using the equation below **(Figure S1C)** where h is the height of the detector, d is the width of the detector, N is the number of pixels along the width of the detector, and p is the amount of time per projection image. Appropriate unit conversions must be used throughout. Plots based on this equation for detectors with 0.3 µm pixels and projection times of 0.1 seconds or 0.01 seconds provide visualization of the rapid growth in imaging rate with detector size **(Figure S1A-B)**. This detector scaling is possible for XRM because of the centimeter-scale depth penetration of X-rays into intact soft tissue biological samples when using beam energies of ~20-50 keV[22] (or higher). Unlike light-sheet fluorescence microscopy (LSFM) or EM methods, XRM nondestructively reconstructs cylindrical tomogram volumes. Since the number of projections needed to reconstruct an image scales linearly with the number of pixels along the horizontal axis and the volume of a tomogram scales (proportional to) cubically, it is theoretically possible to speed up the imaging rate by a square factor as the number of pixels along the axes of the detector grow. It should be noted that detectors do not always possess equal width and height, but I am assuming such dimensions here to simplify the calculations. Similar reasoning applies even if the detector is rectangular.

*Supplemental figures*

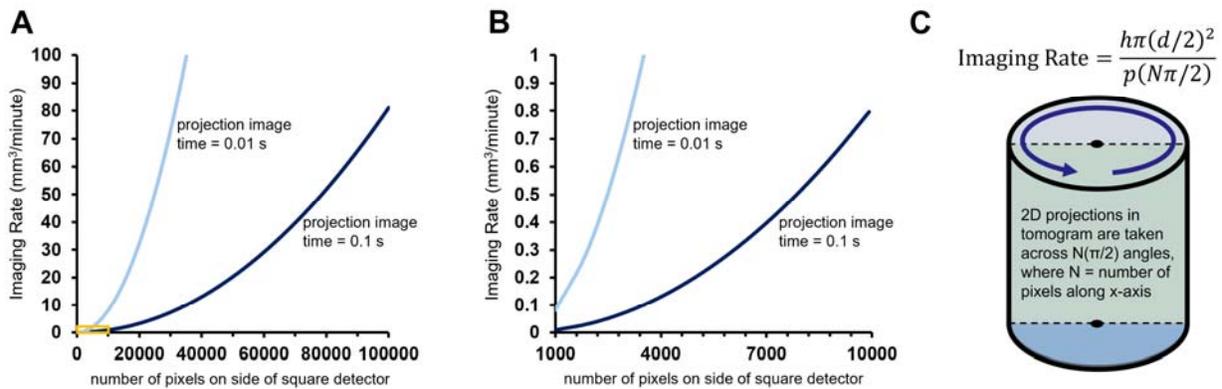

**Figure S1 (A)** Theoretical imaging rate in mm$^3$/minute for XRM systems with square detectors of varying dimensions in terms of the numbers of pixels along a side. Here, a pixel size of 0.3 µm is used. The two curves correspond to imaging rate when projection images each take 0.01 seconds or 0.1 seconds. Similar calculations can be performed for rectangular detectors. **(B)** Zoomed version of the same plot (see orange inset from panel A). **(C)** Imaging rate equation and visualization of a tomogram with description of Crowther's limit.

The equation shown in panel C:
$$\text{Imaging Rate} = \frac{h\pi(d/2)^2}{p(N\pi/2)}$$

2D projections in tomogram are taken across N(π/2) angles, where N = number of pixels along x-axis

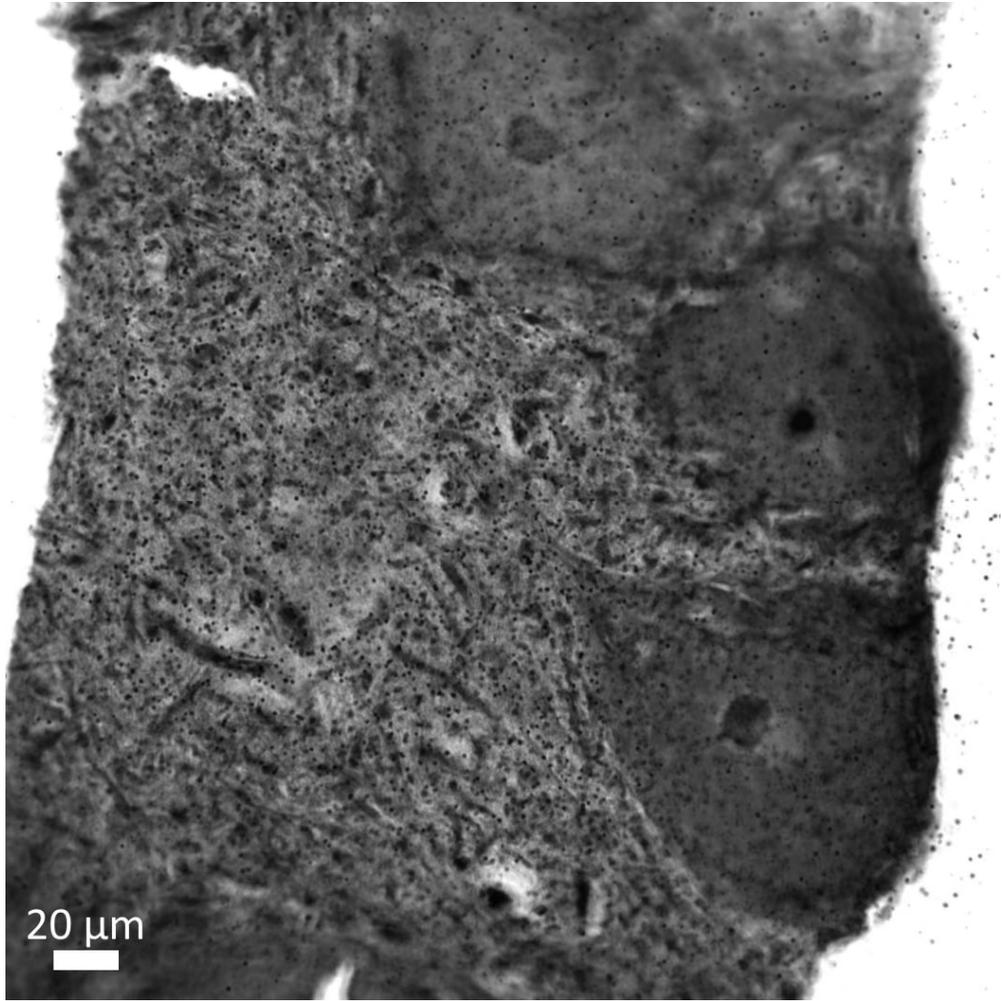

**Figure S2** Black-and-white brightfield microscopy image of DAB-gold sample taken with 40× objective lens.

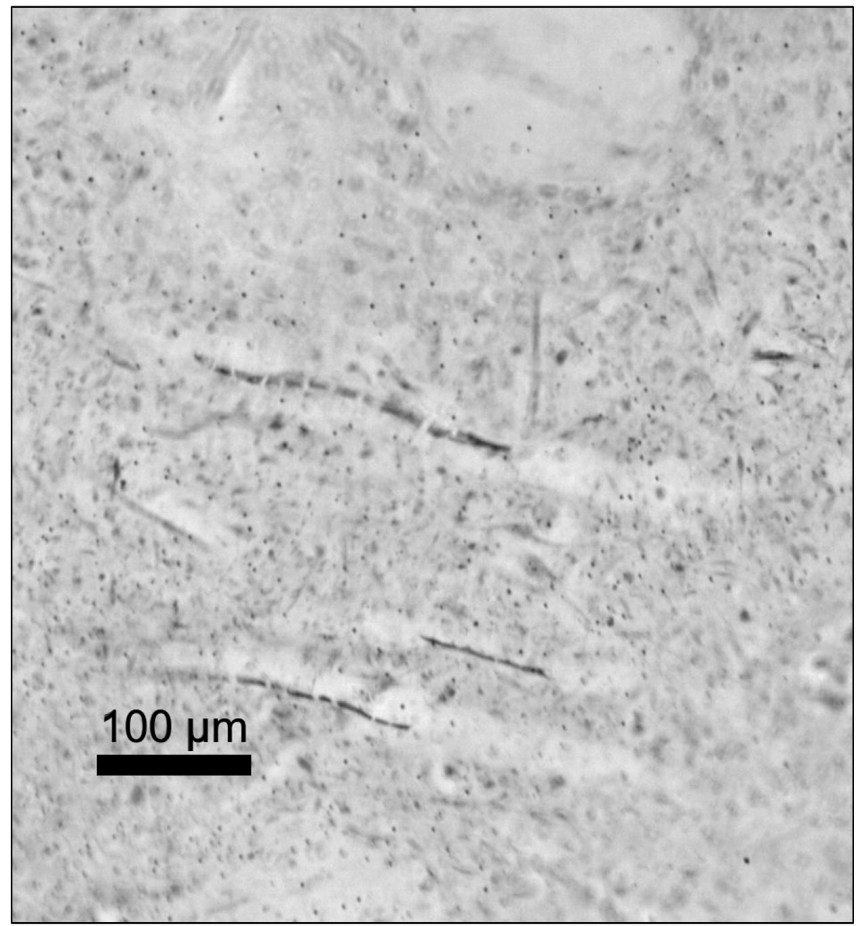 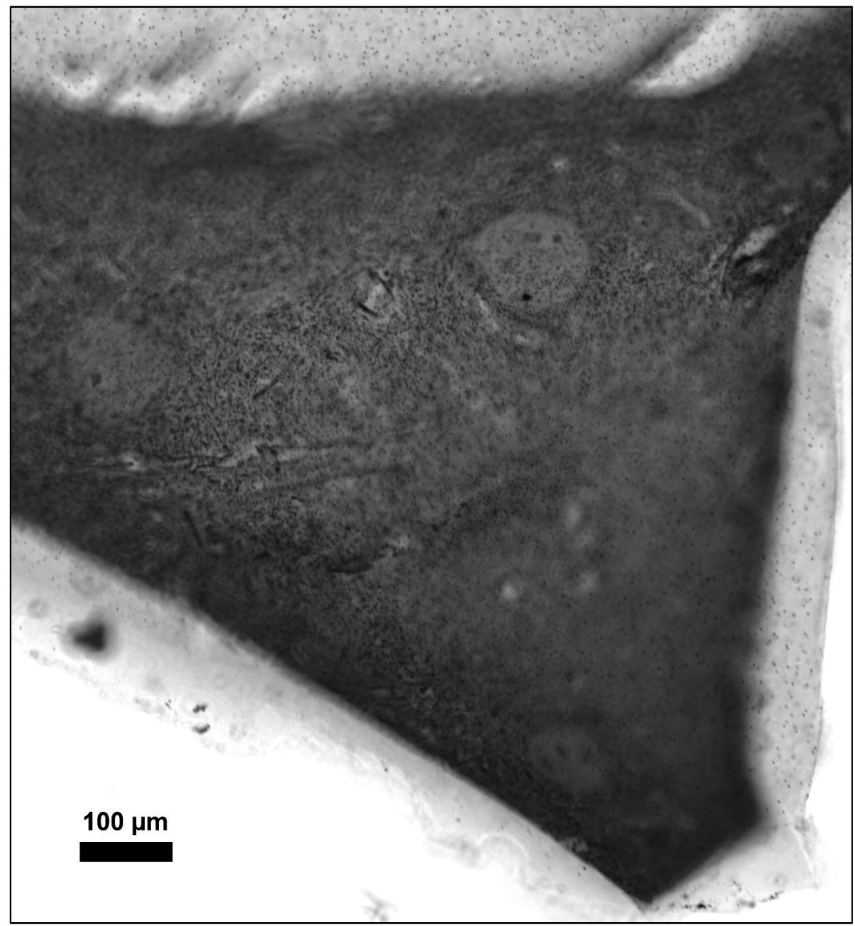 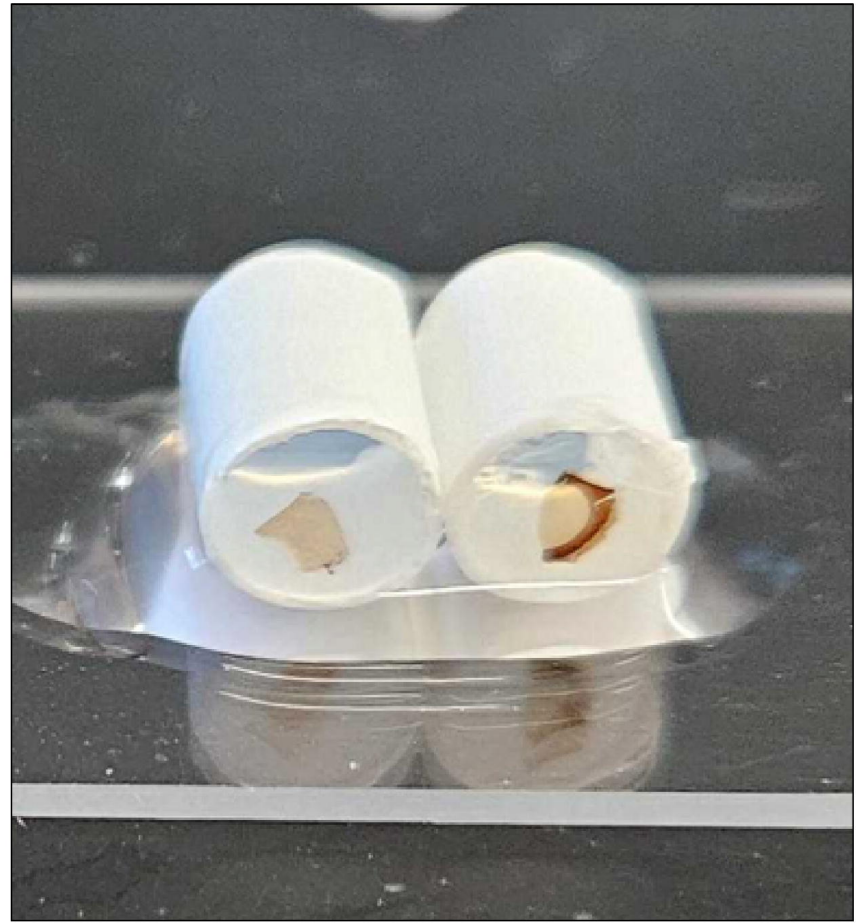

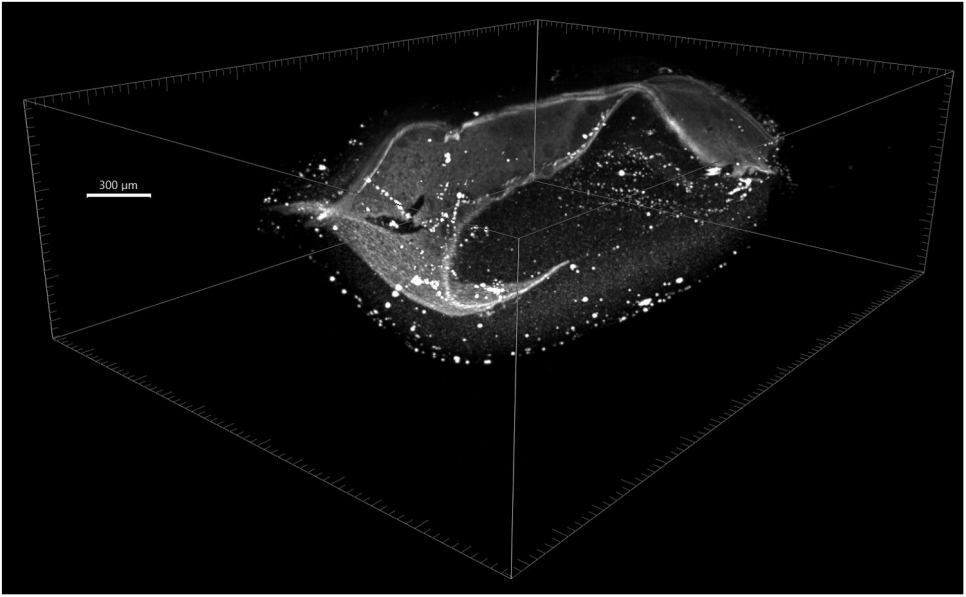 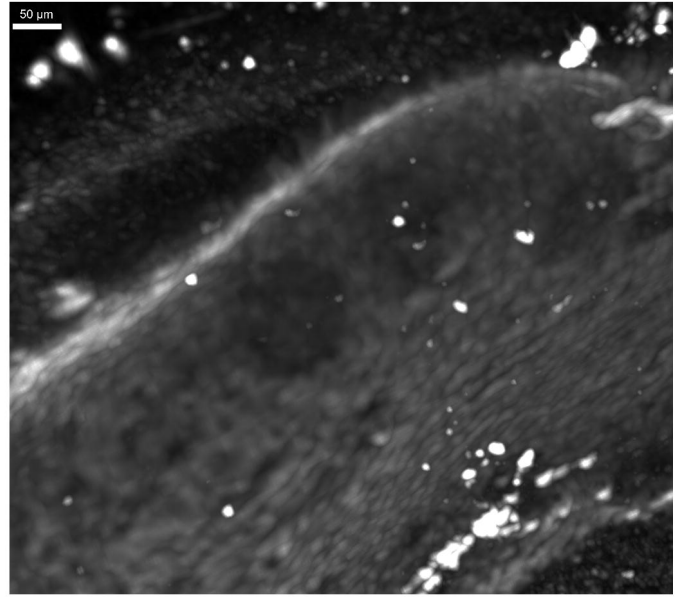 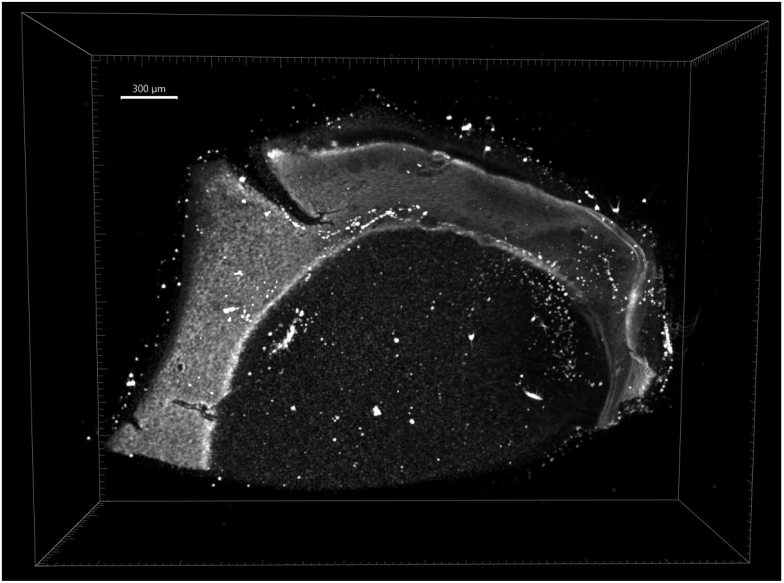 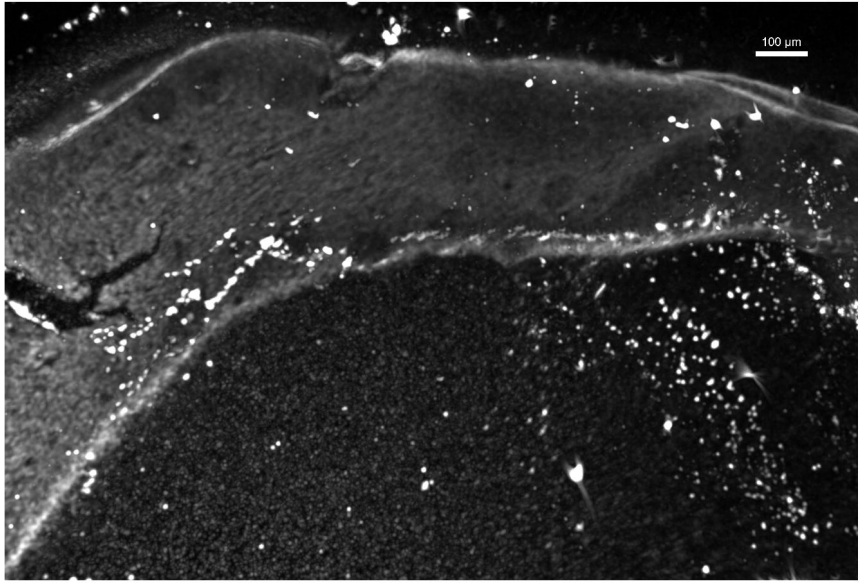

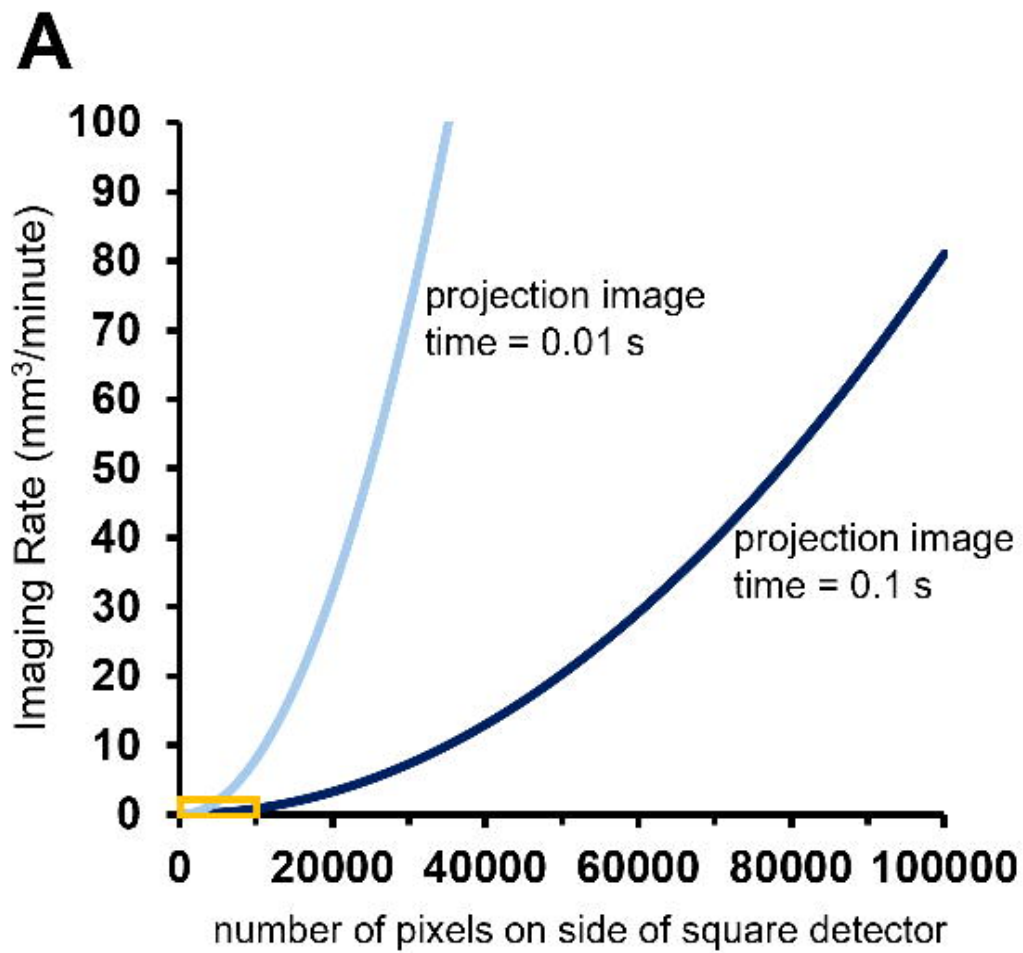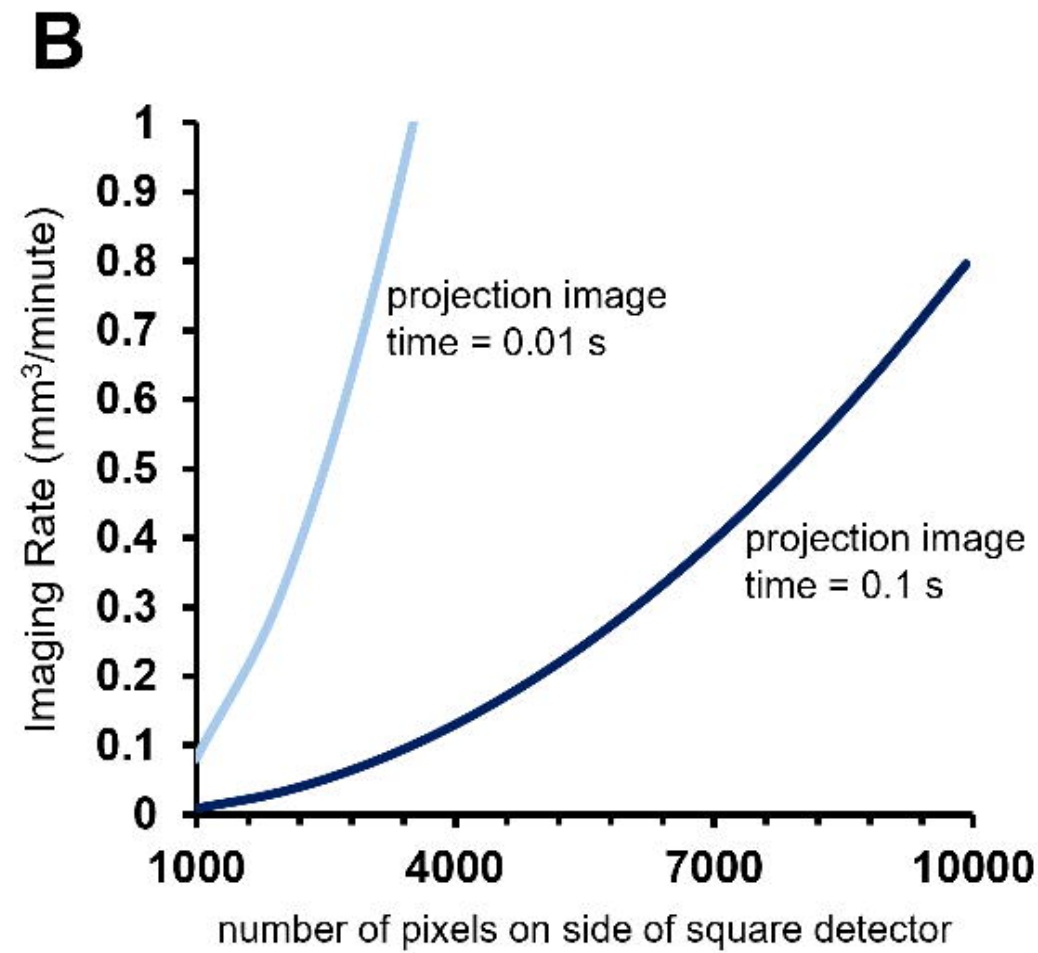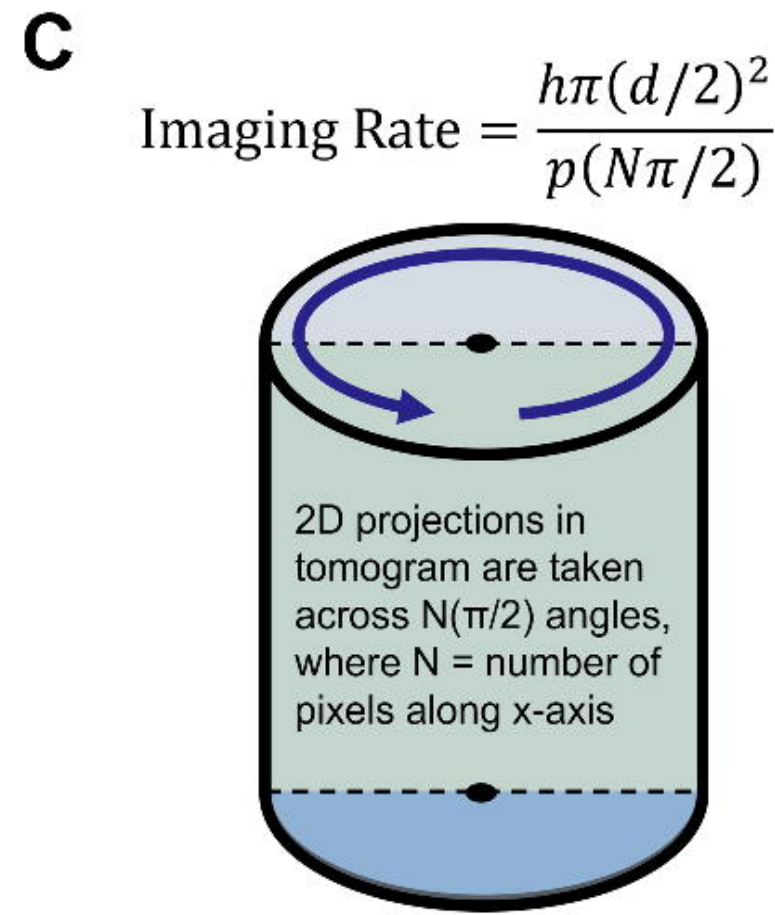

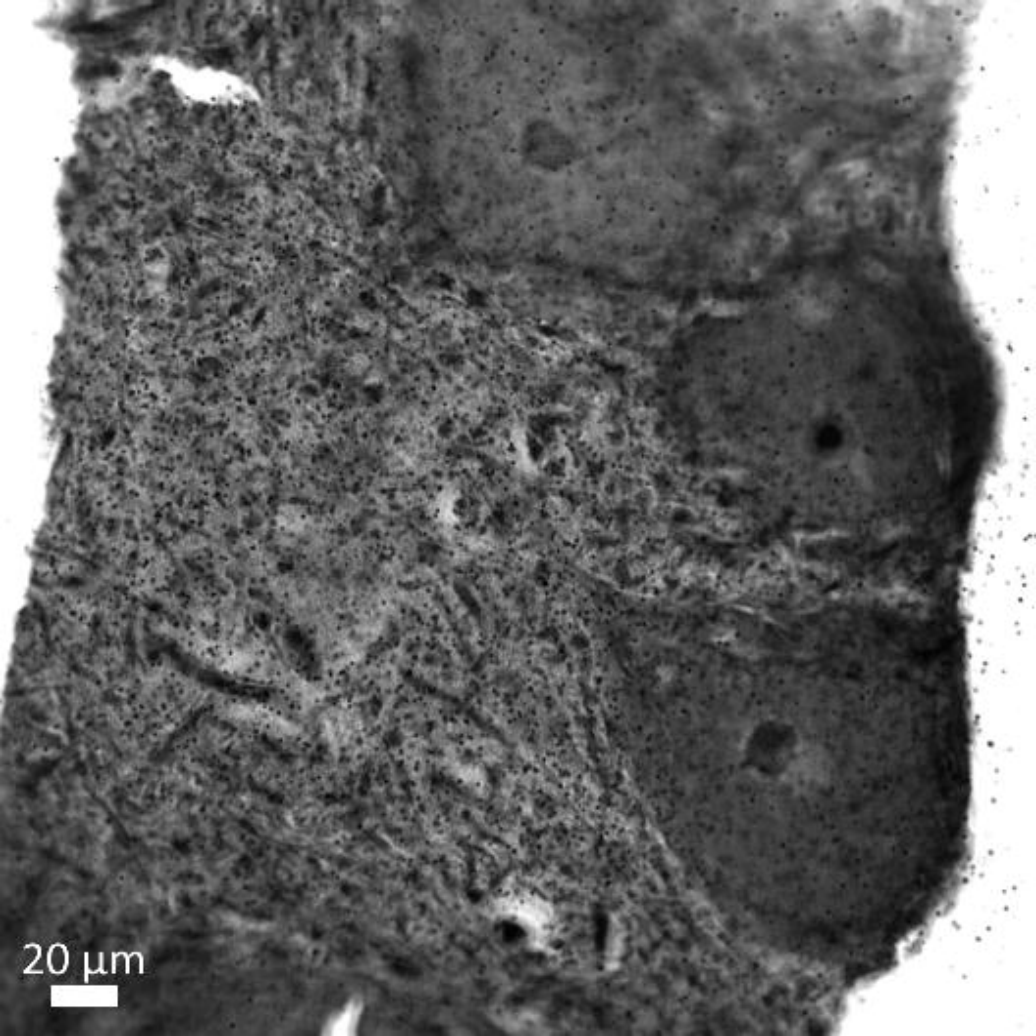
20 μm